\documentclass[journal]{IEEEtran}

\usepackage{amsfonts}
\usepackage{amsmath}
\usepackage{amssymb}
\usepackage{amsthm}
\usepackage{cite}
\usepackage{color}
\usepackage{graphicx}
\usepackage{hyperref}
\usepackage{mathcom}
\usepackage{multirow}
\usepackage{subfig}

\newtheorem{remark}{Remark}
\newtheorem{theorem}{Theorem}

\newcommand{\specialcell}[2][c]{%
  \begin{tabular}[#1]{@{}c@{}}#2\end{tabular}}

\author{
	Jiafan~Yu,~\IEEEmembership{Student~Member,~IEEE,~}%
	Yang~Weng,~\IEEEmembership{Member,~IEEE,~}%
    and~Ram~Rajagopal,~\IEEEmembership{Member,~IEEE}%
	\thanks{J. Yu and R. Rajagopal are with Stanford University, Stanford, CA, 94305 USA e-mail: \{jfy, ramr\}@stanford.edu.}%
	\thanks{Y. Weng is with Arizona State University, Tempe, AZ, 85287 USA e-mail: yang.weng@asu.edu.}%
}

\title{Mapping Rule Estimation for Power Flow Analysis in Distribution Grids}

\begin{document}
\maketitle

\begin{abstract}
The increasing integration of distributed energy resources (DERs) calls for new monitoring and operational planning tools to ensure stability and sustainability in distribution grids. 
One idea is to use existing monitoring tools in transmission grids and some primary distribution grids. 
However, they usually depend on the knowledge of the system model, e.g., the topology and line parameters, which may be unavailable in primary and secondary distribution grids. 
Furthermore, a utility usually has limited modeling ability of active controllers for solar panels as they may belong to a third party like residential customers. 
 To solve the modeling problem in traditional power flow analysis, we propose a support vector regression (SVR) approach to reveal the mapping rules between different variables and recover useful variables based on physical understanding and data mining. 
We illustrate the advantages of using the SVR model over traditional regression method which finds line parameters in distribution grids.
Specifically, the SVR model is robust enough to recover the mapping rules while the regression method fails when 1) there are measurement outliers and missing data, 2) there are active controllers, or 3) measurements are only available at some part of a distribution grid.
We demonstrate the superior performance of our method through extensive numerical validation on different scales of distribution grids.
\end{abstract}

\section{Introduction}
Electric grids are undergoing a profound change. Renewables and other distributed energy resources (DERs) are expected to supply more than $50\%$ of electricity demand by $2050$ in various parts of the world~\cite{EU2050, Jacobson2015}. 
Deep penetration of DERs adds new capabilities and significantly affects the operations of distribution grids. 
In such distribution networks, proper monitoring will be needed for detecting outages~\cite{zhao2013outage}, cyber attacks~\cite{deka2014attacking}, and system failures~\cite{rudin2012machine}. 
In addition to monitoring, operational planning is needed for predicting over-voltage, calculating economic dispatch~\cite{vargas2012real}, and conducting short-term grid controls~\cite{zhang2016agent, jahangiri2013distributed, yeh2012adaptive}.

The power flow equations are the basis for monitoring and planning in distribution grids~\cite{schweppe1970power, andersson2008modelling}.
However, the power flow equations are built through the knowledge of system topology and network parameters. 
Such knowledge is only available in well-maintained primary distribution grids and limited secondary distribution grids.

In many primary and secondary distribution grids, the assumption of complete information does not hold. 
In some secondary distribution grids, only the planned topology and switch locations are known, but real-time changes to the topology can be hard to track. 
Line parameter profiles are inaccurate or even missing. 
Even reconstructing the admittance matrix can be difficult when using distribution management systems (DMS) such as CYME~\cite{CYME}.  
For example, in South California Edison (SCE), they use CYME software to model their distribution grids.
However, the CYME model is only available in a few primary distribution grids. 
Since the CYME model requires all the topology information, line parameter information, as well as the modeling of controllers and loads, it is incapable of modeling many secondary level distribution grids where most of the required information are missing.
Currently, a secondary distribution grid is treated as a single node even if it with DERs such as solar panels.

Future distribution networks will host a variety of active control devices ranging from voltage regulators to inverters for rooftop solar, EV charging and storage. 
These assets are usually independently owned and operated outside of the domain of the DMS. 
The control rules implemented by these devices are unavailable or can be hard to model, making the direct application of power flow analysis difficult and inaccurate~\cite{fan2012probabilistic} even when topology and line parameters are perfectly known. 
Incomplete system information and limited measurements make the system identification problem hard in practice.

The availability of measurements from active devices, line sensors, smart meters, and $\mu$PMUs~\cite{sexauer2013phasor}
is an opportunity to overcome this challenge by designing scalable approaches for system monitoring and analysis relying on new types of data. 
Recent research augments traditional power flow equations by using historical data to initialize state estimators and solvers ~\cite{weng2012search, weng2013historical, weng2016robust}, modifying the current system models~\cite{abdel2013development, wang2012alternative} and proposing novel multi-objective optimization formulations~\cite{sarri2012state}. 
 
In this paper, we focus on building the mapping rules equivalent to the power flow equations in distribution grids.
In particular, we discuss how to design data-driven methods to recover the key relationships in power flow equations: the mapping rules between power injections and voltage phasors.
A distribution grid’s mapping rules are governed by the elements of the admittance matrix when there are unmodeled no active controllers. 
When the accurate measurements of all the historical data, including the set of voltage phasors, real power injections, and reactive power injections, are available at all buses, the admittance matrix could be learned via linear regression using historical data.
The challenges of using the linear regression approach in distribution grids are that the perfect conditions are usually not satisfied.
For example, the parameter estimation approach is not robust against measurement outliers, which are common in distribution grids.
Moreover, the linear regression requires the measurements of all buses.
In distribution grids, usually the measurements at the root (substation or feeder transformers) and the leaves (end users) are available. Other parts of the network have limited measurements for observability.
In this case, a parameter estimation regression model is impossible to build without the measurements at intermediate buses.
Furthermore, since the linear regression model explicitly learn the line parameters, it cannot represent any models beyond the linear relationship.
Therefore, the little flexibility of a linear regression model prevents it capturing the dynamics of any third party-owned controllers.

Finally, the problem of ``inverse mapping'', which recovers the voltage phasor information from the measurements of real and reactive power, are not guaranteed to have a unique solution.
To solve the inverse mapping problem, the information of the topology and line parameters is still a pre-requisite.
With partial measurements and the existence of active controllers, it is hard to recover the full topology and all line parameters through the linear regression model.
Even if we have the information of the topology and line parameters, sometimes, the ``inverse mapping'' problem can be ill-conditioned and do not have a feasible solution.

Therefore, we propose to use a kernel-based support vector regression (SVR)~\cite{friedman2001elements} model to train and represent the mapping rules. 
The insensitive zone of the SVR model and the linear asymptotic behavior of the SVR loss function provide better tolerance over outliers~\cite{smola1997support}.
The kernel trick provides the needed flexibility so that the SVR model can incorporate the power flow equation and incorporate the dynamics of active controllers and handle the situation of incomplete measurements.
Many data-driven models behave like black-box without considering physical law-based models. 
We design the SVR model to represent the based power flow equation exactly when all the measurements are perfectly measured. 
This is achieved by choosing an appropriate kernel in our SVR model. 

The ``inverse mapping'' could be treated as a differentiable mapping.
The new mapping is a function of real and reactive power, with output voltage magnitudes.
Thus, locally, it can be approximated by a polynomial function of real and reactive power through Taylor expansion.
The flexibility of the SVR model with kernel trick provides an accurate approximation for the locally expanded polynomial function.

Furthermore, SVR can be computed very efficiently using interior point methods and distributed computing~\cite{moulin2004support, menon2009large, ho2012large} and many different kernels can be utilized depending on the applications~\cite{bishop2006pattern}.

Some other black box-like data-driven models, such as neural networks (NN), could be also used for the purpose.
However, NN does not guarantee the exact representation of the traditional physical law-based model.  
For example, NN usually requires more data than SVR and is possible to overfit. 
It works best for highly nonlinear system such as image recognition and natural language processing.
In our situation, the system is still governed by physical laws, the SVR model can identify the physical understanding behind the data, but NN cannot.

We test the proposed SVR model for estimating both the forward and inverse mapping rules 
between voltage phasors and power injections on different scales of distribution grids.
For example, we use IEEE 8, 123 bus, and systems with bus number between 8 and 123.
We also compare the SVR model with traditional parameter learning-based regression.
The results reveal that the SVR model outperforms traditional models, especially for the cases of partial measurements, system with active controllers and measurements with outliers.
The satisfactory results reveals that we can use the SVR-based mapping rule estimation as the equivalence of the traditional physical law-based power flow equations in distribution grids with renewables.

The rest of the paper is organized as follows:
Section~\ref{sec:a} reviews the power flow analysis and defines the problem of learning mapping rules for distribution networks.
Section~\ref{sec:b} shows how the mapping rule learning can be represented as a SVR problem and how to embed power systems physical understanding into the SVR model.
Section~\ref{sec:c} illustrate the advantages of SVR model over traditional parameter learning-based regression model.
Section~\ref{sec:d} analyzes experimental results on different distribution grids and compares to the traditional physical law-based power flow equations.
Section~\ref{sec:e} concludes the paper.

\subsection*{Notation}
We use lower case English and Greek letters, such as $p, \beta$ to denote scalars and scalar functions, use lower case bold English and Greek letters, such as $\mathbf{a}$, $\phiv$ to denote vectors and vector functions.
We use upper case English letters, such as $G$ to denote matrices.
We use a comma (,) to denote horizontal concatenation of vectors, and we use a semicolon (;) to denote vertical concatenation of vectors.
For example, $[x_1, x_2] \in \mathbb{R}^{1\times 2}$ is a row vector, and $[x_1; x_2] \in \mathbb{R}^{2\times 1}$ is a column vector.

\section{Problem Motivation and Formulation}\label{sec:a}
For traditional grid monitoring and planning tools, the physical power flow mappings serve as the basis~\cite{lesieutre2011examining}:
\begin{subequations}\label{eqn:a}
	\begin{align}
		p_i = & \sum_{k=1}^n |v_i||v_k| (g_{ik}\cos \theta_{ik} + b_{ik}\sin \theta_{ik}),\\
		q_i = & \sum_{k=1}^n |v_i||v_k| (g_{ik}\sin \theta_{ik} - b_{ik}\cos \theta_{ik}),
	\end{align}
\end{subequations}
where $i = 1, \cdots, n$. 
$p_i$ and $q_i$ are the real and reactive power injections at bus $i$, $(g_{ik} + j \cdot b_{ik})$ is the $(i, k)$-th element in the admittance matrix $Y=G+j \cdot B$, where $j$ is the imaginary unit. 
$|v_i|$ is the voltage magnitude at bus $i$ and $\theta_{ik}$ is the phase angle difference between bus $i$ and bus $k$. 

To use the kernel-based analysis in the following content, we use the rectangular coordinate to represent the voltage phasor because the rectangular coordinate representation simplifies the trigonometric functions to polynomial functions.
By defining
\[
	u_i = |v_i| \cos \theta_i,\quad w_i = |v_i|\sin \theta_i,
\]
where $u_i$ and $w_i$ are the real and imaginary components of the voltage phasor, the physical law-based power flow mappings~\eqref{eqn:a} can be also expressed as functions of $u_i$ and $w_i$:
\begin{subequations}\label{eqn:b}
	\begin{align}
		p_i = & \sum_{k=1}^n (u_iu_kg_{ik} + w_iw_kg_{ik} + w_iu_kb_{ik} - u_iw_kb_{ik}),\\
		q_i = & \sum_{k=1}^n(w_iu_kg_{ik} - u_iw_kg_{ik} - u_iu_kb_{ik} - w_iw_kb_{ik}).
	\end{align}
\end{subequations}
Furthermore, we denote 
\[
\uv = [u_1; \cdots; u_n], \quad \wv = [w_1; \cdots; w_n], \quad \xv = [\uv; \wv].
\] 
Then, the inherent power flow mappings~\eqref{eqn:b} can be abstractly represented as
\begin{subequations}\label{eqn:c}
	\begin{align}
		p_i = & f_{p_i}(\xv),\\
		q_i = & f_{q_i}(\xv).
	\end{align}
\end{subequations}

Traditionally, the power flow mappings $f_{p_i}$ and $f_{q_i}$ are determined by the system topology and line parameters. 
However, in distribution grids, the physical law-based representation of power flow mappings may be unavailable because of inaccurate topology information and missing line parameters. 
To solve the problem, we observe increasing data availability in distribution grids.
Therefore, we propose to directly represent the power flow mappings from one set of measurements to another solely based on historical measurements of $\xv_t$ and ${p_i}_t$ (or ${q_i}_t$), $t=1, \cdots, T$~\cite{friedman2001elements}.

\subsection{Representing the Power Flow Mappings using Inner-Product}
\begin{figure*}[!t]
	\centering
	\includegraphics[width=0.9\textwidth]{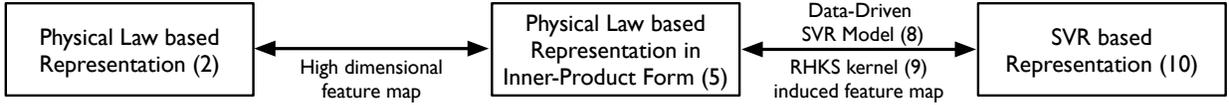}
	\caption{Illustration of Representation Transformation.}
	\label{fig:a}
\end{figure*}
When estimating the power flow mappings, $f$ can be expressed in a different form than \eqref{eqn:b} to emphasize the unknown coefficients $g_{ik}$ and $b_{ik}$:
\begin{subequations}\label{eqn:d}
	\begin{align}
		p_i = & \sum_{k=1}^n g_{ik}\left(u_iu_k + w_iw_k\right) + b_{ik}\left(w_iu_k - u_iw_k\right),\\
		q_i = & \sum_{k=1}^n g_{ik}\left(w_iu_k - u_iw_k\right) - b_{ik}\left( u_iu_k + w_iw_k\right).
	\end{align}
\end{subequations} 
Subsequently, the power flow mappings~\eqref{eqn:d} could be treated as the inner-product between the vector $[\gv_i; \bv_i]$ and a feature mapping $\phiv(\cdot)$ of the state vector $[\uv; \wv]$:
\begin{subequations}\label{eqn:e}
	\begin{align}
		p_i = & \left<[\gv_i; \bv_i], \phiv_{p_i}([\uv; \wv])\right>,\\
		q_i = & \left<[\gv_i; \bv_i], \phiv_{q_i}([\uv; \wv])\right>,
	\end{align}
\end{subequations}
where $<\cdot, \cdot>$ represents the inner product of two vectors,
$\gv_i = [g_{i1}; \cdots; g_{in}]$, $\bv_i = [b_{i1}; \cdots; b_{in}].$
In other words, if we map the state vector $\xv = [\uv; \wv]$ to a higher dimensional space, the power flow mapping becomes a linear function between $p_i$ and $\phiv_{p_i}(\xv)$ with parameters $[\gv_i; \bv_i]$.
After compactly denoting 1) the output as $y$, 2) the system model parameter as $\betav$, and 3) the state vector as $\xv$, the power flow mapping could be expressed as:
\begin{equation}\label{eqn:f}
y = \left<\betav_y, \phiv_y (\xv)\right>,
\end{equation}
where $y = p_i$ (or $y= q_i$), $\betav_{p_i} = \betav_{q_i} = [\gv_i; \bv_i]$.

\section{Support Vector Regression for Power Flow}\label{sec:b}
\subsection{Estimating Model Parameter via Linear Regression}
The power flow mapping~\eqref{eqn:f} is linear with respect to the system parameters $\betav_y$.
A straightforward approach to find the mapping is to estimate the physical model parameter $\betav_y$ directly through linear regression based on historical data points $(\xv_t, y_t)$, $t=1, \cdots, T$.
By defining
\[
\Phi_y := \left[\begin{array}{c}\phiv_y\left(\xv_1\right)^T\\ \vdots \\ \phiv_y\left(\xv_T\right)^T\end{array}\right],
\]
the least-square estimation of $\betav_y$ is:
\begin{equation}
\widehat{\betav}_{y,~\text{LS}} = \left(\Phi_y^T \Phi_y\right)^{-1} \Phi_y^T \yv.
\end{equation}

\subsection{Formulating the Power Flow Mapping Estimation Problem using the SVR Model}
\begin{figure}[!t]
	\centering
	\includegraphics[width=0.45\textwidth]{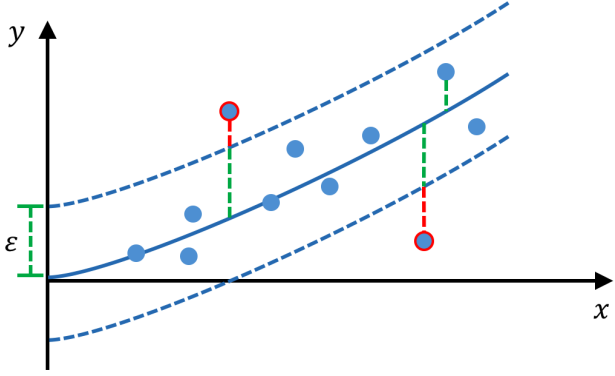}
	\caption{Illustration of SVR.
	The dots are training data points.
	The $x$-axis represents the feature, and the $y$-axis represents the response.
	$\epsilon$ is the range of no-penalty tube. There are two data samples outside the $\epsilon$-insensitive zone. The red vertical dashed lines indicate the associated penalties.}
\label{fig:b}
\end{figure}

Besides the linear regression approach, the inner-product representation of the power flow mappings naturally forms the basis of a support vector regression (SVR) model~\cite{scholkopf2002learning} to estimate the mappings in~\eqref{eqn:f}:
\begin{subequations}\label{eqn:g}
	\begin{align}
		\begin{aligned}
			&\underset{\betav, \xiv, \xiv^\star, b}
			{\text{minimize}} \quad \frac{1}{2}\left\|\betav\right\|^2 + C\sum_{\tau=1}^T \left(\xi_t + \xi_t^\star \right)\label{eqn:h} \quad \quad \quad~
		\end{aligned}
		\\
		\begin{aligned}
			&\text{subject to} & & y_t - <\betav, \phiv_y\lb\xv_t\rb> - b  \leq \epsilon + \xi_t,\label{eqn:i}
		\end{aligned}\\
		\begin{aligned}
			& & & <\betav, \phiv_y\lb\xv_t\rb> + b  - y_t \leq \epsilon + \xi_t^\star,\label{eqn:j}
		\end{aligned}\\
		\begin{aligned}
			& & &\xi_t, \xi_t^\star \geq 0,\quad \quad \quad \quad \quad \quad \quad \quad \quad~
		\end{aligned}
	\end{align}
\end{subequations}
where $t = 1, \cdots, T$, are $T$ samples from historical data.
In particular, the inequality constraints~\eqref{eqn:i} and~\eqref{eqn:j} set zero penalty for training data samples located in the $\epsilon$-insensitive zone, in which the data samples contribute no error to the regression fit, or $\xi_t =0$ and $\xi_t^\star=0$.
Only the training data samples outside the $\epsilon$-insensitive zone determine the optimal fitting.
These data samples are called support vectors.
An illustration of a typical SVR fit is shown in Fig.~\ref{fig:b}. 

\subsection{SVR Power Flow}\label{subsec:b}
The SVR optimization in~\eqref{eqn:g} is in general difficult to solve due to the large number of constraints and the dimension of the feature mapping  $\phiv(\xv)$. 
However, special choices of feature mappings lead to a simple representation of the solutions for the SVR regression. 
These feature mappings satisfy the kernel trick property:
\begin{equation}
K(\xv_1, \xv_2):= \left<\phiv(\xv_1), \phiv(\xv_2)\right> = h\left(\left<\xv_1, \xv_2\right>\right),
\end{equation}
where the inner-product between $\phiv(\xv_1)$ and $\phiv(\xv_2)$ is a scalar function of the inner-product between $\xv_1$ and $\xv_2$, and $h(\cdot)$ is a scalar function~\cite{bishop2006pattern}.
The space of such feature mappings satisfying this property is the reproducing Hilbert kernel space (RHKS).

By choosing a feature mapping $\phiv(\cdot)$ in RHKS, we can avoid directly calculating the feature mapping and estimating the topology and line parameters explicitly in the intermediate step.
Instead, the kernel automatically helps mapping the data to a proper higher dimension space.
To solve the optimization problem~\eqref{eqn:g}, we only need to calculate the inner-product between different training data samples:
$K(\xv_{t_1}, \xv_{t_2}) = h\left(\left<\xv_{t_1}, \xv_{t_2}\right>\right)$.

Furthermore, the solution of~\eqref{eqn:g} does not directly provide the optimal model parameter $\betav^\star$.
Instead, the solution of~\eqref{eqn:g} is given by an optimal set of parameters $\alpha_t^\star$, $t=1, \cdots, T$.
Therefore, the power flow mappings~\eqref{eqn:b} could be represented as the linear combination of the kernel product between a state $\xv$ and the historical data $\xv_1, \cdots, \xv_T$, parameterized by $\alphav^\star$: 
\begin{equation}\label{eqn:k}
y = f_y^\star(\xv) = \sum_{t=1}^T \alpha_t^\star K(\xv, \xv_t).
\end{equation}
The $\alpha_t^\star$ is nonzero only when $\xv_t$ is a support vector.
This fact makes the SVR-based representation of power flow mapping sparse and easy to compute.

As an illustration, Fig.~\ref{fig:a} summarizes the transformation from the physical law-based representation to the historical data-driven SVR-based representation of the power flow mappings. 
\begin{remark}
The physical law-based representation~\eqref{eqn:e} and the SVR-based representation~\eqref{eqn:k} of power flow mappings are both defined using inner-products.
However, these two representations have fundamental differences.
The parameters of the physical law-based representation are line parameters $[\gv; \bv]$, of which the dimension is proportional to the size of the distribution grid. 
Moreover, to apply the mapping representation, we must explicitly mapping the state $\xv$ to a higher dimensional space $\phiv(\xv)$ and conduct inner-products.
In contrast, the parameters of the SVR-based representation are solely the historical data sample $X$ and the associated Lagrangian multipliers $\alphav^\star$, of which the dimension is the number of support vectors, independent from the size of the distribution grids.
Furthermore, to apply the SVR-based mapping representation, we only need to conduct the kernel inner-product between the state $\xv$ and historical data samples, without explicitly mapping the state to higher dimensional space.
This is specially useful in distribution grids where the data is abundant but a complete physical model is missing. 
\end{remark}

\subsection{Choosing Tuning Parameters for SVR Power Flow}
Cross-validation is typically utilized in SVR to choose the tuning parameters $C$ and $\epsilon$ in~\eqref{eqn:g} ~\cite{hastie09}. 
This enables the method to increase robustness towards noise and outliers in the data. It also ensures that SVR has good predictive performance. 

The suggested approach for SVR-based Power Flow is to utilize k-fold cross-validation ~\cite{hastie09} (typically $k=5$) with the training data to select the optimal choices of $C$ and $\epsilon$, with $k-1$ blocks of data used to train the model and one block utilized to assess validation performance and select tuning parameters.  The SVR performance is then assessed in a separate data set. The choice of parameters determines the sparsity of $\alpha_t$ in the kernel representation. 

\subsection{Generalizing Power Flow SVR: Inverse Mappings}\label{subsec:a}
In many applications of power flow analysis, we are interested in recovering voltage magnitude and phase angle information from the measurements of real and reactive power. Typically, a calibrated power flow model is utilized and solved. Power flow solutions are not guaranteed to be unique, and in some instances, the problem can be ill-conditioned. 
Additionally, the system might not be fully observed preventing learning of an accurate model in the absence of topology information and relatively accurate line parameter data.
\eqref{eqn:g} instead enables learning an  \emph{inverse mapping} of voltage magnitude as a function of power from historical data.  
The inverse power flow is a differentiable mapping as a function of real and reactive power. 
Thus, locally, $|v_i|$, the voltage magnitude for bus $i$, can be approximated by a polynomial function of ${[\pv;\qv]}$. 
Setting  $\xv = [\pv;\qv]$ and utilizing the polynomial kernel produce approximations that can achieve arbitrary accuracy with respect to the Taylor expansion of the inverse mapping. 
\section{Advantages of Using SVR Representation over Regression}\label{sec:c}
\subsection{Connection between SVR Model and Physical Model}
When the historical measurements at all buses are fully observable and there are no measurement errors, we have the following theorem proving that the SVR-based representation of power flow mappings can exactly recover the physical law-based representation:
\begin{theorem}\label{thm:a}
The physical law-based power flow mappings~\eqref{eqn:b} can be exactly represented by choosing the quadratic kernel 
\begin{equation}\label{eqn:l}
K(\xv_1, \xv_2) = \left(\left<\xv_1, \xv_2\right> + c\right)^2 = \left(\xv_1^T \xv_2 + c\right)^2.
\end{equation}
\end{theorem}
\begin{proof}
First, the quadratic kernel is in the reproducing Hilbert kernel space (RHKS).
The feature mapping corresponding to the quadratic kernel~\eqref{eqn:l} is
\begin{equation}\label{eqn:m}
\begin{aligned}
\phiv(\xv) &= [x_1^2, \cdots, x_m^2, \sqrt{2} x_1x_2, \cdots, \sqrt{2} x_1x_m,\\
&\sqrt{2} x_2x_3, \cdots, \sqrt{2}x_{m-1}x_m, \sqrt{2}cx_1, \cdots, cx_m, c].
\end{aligned}
\end{equation}
Second, we can constructively build a $\betav$ such that the inner-product between $\betav$ and the quadratic feature mapping $\phiv(\xv)$ exactly recover the power flow mapping for $p_i$.
Given $\xv = [\uv; \wv]$ and the feature mapping $\phiv(\cdot)$ in~\eqref{eqn:m}, we define $\betav^\star$ as following:
\begin{equation}\label{eqn:n}
    \beta^\star_j = 
\begin{cases}
    g_{ii},& \text{if }\phiv(\xv)_j = u_i^2\text{ or }\phiv(\xv)_j = w_i^2,\\
    \frac{1}{\sqrt{2}} g_{ik},& \text{if }\phiv(\xv)_j = \sqrt{2}u_iu_k\text{ or } \sqrt{2}w_iw_k, i \neq k,\\
    \frac{1}{\sqrt{2}} b_{ik},& \text{if }\phiv(\xv)_j = \sqrt{2}w_iu_k,\\  
    -\frac{1}{\sqrt{2}}b_{ik},& \text{if }\phiv(\xv)_j = \sqrt{2}u_iw_k,\\        
    0, & \text{otherwise}.
\end{cases}
\end{equation}
With the definition of $\phiv(\xv)$ in~\eqref{eqn:m} and $\betav$ in~\eqref{eqn:n}, the inner-product between $\betav$ and $\phiv(\xv)$ is exactly the physical law-based mapping from $\xv$ to $p_i$:
\begin{equation}
\begin{aligned}
p_i &= \sum_{k=1}^n g_{ik}\left(u_iu_k + w_iw_k\right) + b_{ik}\left(w_iu_k - u_iw_k\right)\\
& = \left<\betav^\star, \phiv(\xv)\right>.
\end{aligned}
\end{equation}
\end{proof}

\subsection{Robustness of SVR Model against Outliers}
The parameter learning-based regression model only works good if the data is outlier-free.
This is because the loss function of a linear regression is a quadratic function.
On finite samples, the squared-error loss places much more emphasis on observations with large absolute residuals $|y_t - f(\xv_t)|$ during the fitting process.
It is thus far less robust, and its performance severely degrades for grossly mis-measured $y$-values (``outliers'')~\cite{friedman2001elements}.

The least absolute value deviation estimation (LAD)~\cite{bloomfield1980least} replaces the quadratic loss function by absolute value loss function which provides a more robust criteria.
However, the LAD model cannot guarantee a unique solution~\cite{branham1982alternatives}.
There are possibly multiple solutions achieving the minimal loss function value.

\begin{figure}[!t]
	\centering
	\includegraphics[width=0.45\textwidth]{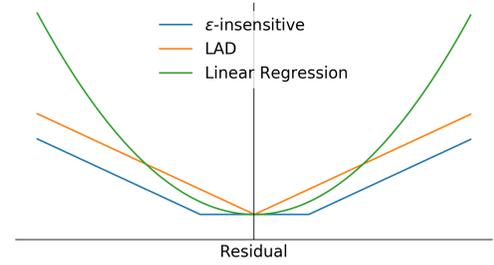}
	\caption{Illustration of loss functions for linear regression, least absolute value deviation estimation, and support vector regression.}
	\label{fig:c}
\end{figure}

The SVR model resolves the drawbacks of the traditional regression and the least absolute value deviation estimation.
First, the asymptotic behavior of the $\epsilon$-insensitive loss function is linear, which is less sensitive to large absolute residuals.
Furthermore, the regularization of the parameter $\betav$ in the loss function eliminates the possibility of multiple optimal solutions, which makes the SVR model much more stable than the LAD model.
We compare the different loss functions of linear regression model, LAD model and SVR model in Fig.~\ref{fig:c}.
\begin{figure}[!t]
	\centering
	\includegraphics[width=0.45\textwidth]{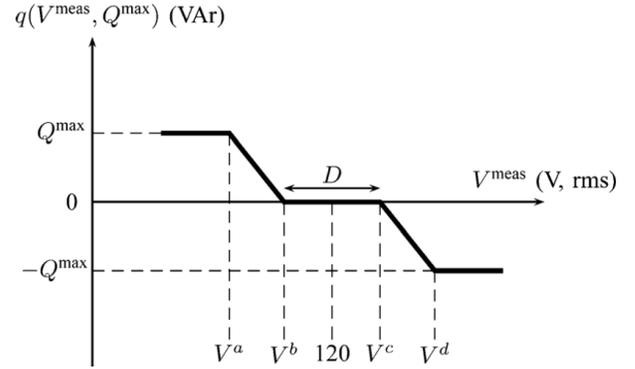}
	\caption{Droop control function for a United States-based system.}
	\label{fig:d}
\end{figure}

\subsection{Flexibility of SVR Model for Active Controller Modeling}
Increasing penetration of DER adds a variety of active controllers whose control algorithms and device models might be unavailable to the power monitoring systems at the utility. 
One common kind of active controllers is a capacitor power bank for volt/var control to maintain a stable voltage.
Fig.~\ref{fig:d} illustrates a droop control function where the additional reactive power injection is determined by the magnitude of voltage.
These active controllers affect the physical law-based power by adding unmeasured power injections to distribution grids.
For instance, if bus $i$ in an $n$-bus distribution grid is equipped with a reactive power bank, where injected reactive power follows voltage variation, $q_i' = h(\vv, \thetav)$, the modified power flow equation at bus $i$ changes to:
\begin{subequations}\label{eqn:o}
	\begin{align}
		q_i = & \sum_{k=1}^n |v_i||v_k| (g_{ik}\sin \theta_{ik} - b_{ik}\cos \theta_{ik}) - q_i',
	\end{align}
\end{subequations}
where $q_i'$ is the additional power injection from the reactive power bank, and $h(\cdot)$ is the control policy.
If $q_i'$ is omitted from the model, an incorrect mapping will be obtained. 

\begin{figure}[!t]
	\centering
	\includegraphics[width=0.45\textwidth]{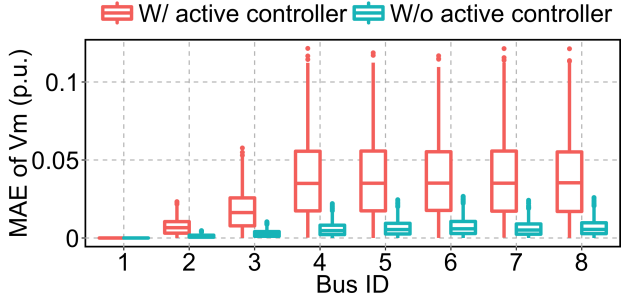}
	\caption{Performance of traditional power flow solver (Newton-Raphson method) for voltage magnitude estimation, with distributed active controller at bus $4$, given accurate topology and line parameters.}
	\label{fig:e}
\end{figure}

As an illustration, we add a reactive power controller at bus $4$ for the IEEE $8$-bus distribution grid and assume topology and line parameters are known.
Fig.~\ref{fig:e} shows the significant mean absolute error (MAE) appearing in the traditional power flow analysis when an active device is added but not modeled.

The flexibility of SVR model also provides a practical approach to represent the third-party owned distributed controllers' model once the control algorithm is a differentiable mapping as a function of real and reactive power.

\subsection{Flexibility of SVR Model for Partial Observable Distribution Grids}
\begin{figure}[!t]
	\centering
	\includegraphics[width=0.45\textwidth]{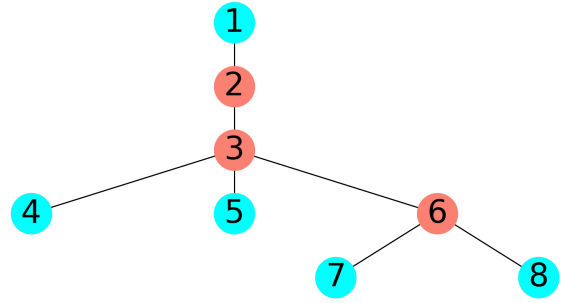}
	\caption{Illustration of a partially observed distribution grid.
	Power is injected from bus 1, and is consumed at bus 4, 5, 7, and 8.
	The measurements are unavailable at bus 2, 3, and 6 which are marked red.}
	\label{fig:f}
\end{figure}

In many distribution grids, the measurements are only available at the root level where the substation/feeder transformers are located, and the leaves level where the residential loads and distributed energy resources are located.
In the intermediate level buses, no measurements are available.
In this case, neither the system is fully observable, nor the regression model can provide the correct line parameters.
However, due to the flexibility of the kernel-based SVR model, we can still use the measurements at the available buses as the input to have an accurate power flow representation between the partially measured voltages and power injections at root/leaves.
Fig.~\ref{fig:f} shows an example of a partially-observed distribution grid, where the measurements on red nodes are unavailable.
Since the traditional regression model requires all measurements to calculate $v_i v_j \cos (\theta_i - \theta_j)$ and $v_i v_j \sin (\theta_i - \theta_j)$,
in this case, the regression model can never get the correct line parameters.
If the hidden nodes are without source, we can prove that there exists an equivalent admittance matrix.
If there are power injections at hidden buses, the regression-based model fails.
However, the power injections are still determined by the voltages of the available nodes (bus 1, 4, 5, 7, and 8), the flexibility of the SVR model guarantees that it can still capture the mapping rules.
While the physical law-based model does not have a meaningful interpretation for partially-observable grids, the SVR representation model still captures the temporal relationship between the mapping rules and the historical data.

\section{Experimental Results}\label{sec:d}
\subsection{Experiment Setup}
We test our data-driven power flow approach on a variety of settings and real-world data sets. 
We use $8, 16, 32, 64, 96, 123$-bus test feeders, as well as two Southern California Edison (SCE) distribution networks with different shapes.
Here, the $16, 32, 64, 96$-bus systems are extracted from the IEEE $123$-bus system.
The bus power injection data is from primary distribution grids of Southern California Edison (SCE) and secondary distribution grids of Pacific Gas and  Electric (PG\&E).
The real and reactive power injection data are from small and medium business or aggregate of several residential homes. The sampling frequency is one hour.
The SCE data set's period is from January $1$, $2015$, to December $31$, $2015$, and the PG\&E data set's period is from August $1$, $2010$, to July $1$, $2011$.
For IEEE standard test feeders, we run power flow using Matpower package~\cite{josz2016ac} to obtain the associated voltage magnitudes and phase angles at each bus.
We use MATPOWER software only to implement the Newton-Raphson iteration.
All the parameters are built through either IEEE standard distribution grid models or South California Edison real data. 
For SCE distribution networks, the voltage phase angle information are available on some buses.
In our experiments, the topology and line parameter information from IEEE standard case files are only used for data preparation to build the relationship between power injections and voltage phasors.
In all evaluation steps, we assume that the topology information and the line parameters are unavailable.
Finally, noises are added to check the robustness of the proposed approach.

In particular, we compare the regression-based approach and our proposed SVR approach for three common scenarios in distribution grids: 1) with the existence of outliers, 2) with unknown volt/var controllers, and 3) with partially observable distribution grids.
All these scenarios are tested on different scales of distribution grids including radial networks and mesh networks.

\subsection{Data Collection}
Currently, South California Edison (SCE) is regularly collecting different types of time series data from their distribution grids, including the voltage and power from the root nodes as well as the nodes behind the root.
Currently, we are building the VADER data infrastructure which has the capability of handling heterogeneous data and post-processing them.
We have different data extractors to acquire data, and have a unified data distributor to clean and sort the acquired data to a Cassandra database, which serves as the input source for the proposed mapping rule estimation.

\subsection{Effectiveness of SVR Model for Forward Mapping}
We prove that the power flow equations can be represented exactly by the proposed SVR model with the 2nd order polynomial kernel in~\eqref{eqn:k}, when we choose the rectangular coordinate representation of the state vector.
For forward mapping, we build the proposed SVR model with 2nd order polynomial kernel and other RHKS kernels, as well as the parameter learning-based regression model and an averaging model.
In particular, the input of all these models is the voltage phasors at all buses, and the output is the power injection at one certain bus.
The raw input and output variables are all the same for different models.
Then, for parameter learning-based regression model, raw inputs are reformulated following~\eqref{eqn:b}.
For SVR models, raw inputs are only reformulated to rectangular coordinates with reference phase angle 45$^\circ$.
For performance comparison, we use six weeks' hourly sampled historical data of voltage phasors from all buses and power injections at some buses to train all these models.
Random Gaussian measurement error with $1\%$ relative standard deviation are added to training data.
$2\%$ of training data samples are modified to outliers.
Another three weeks' data is used for testing the performance.
No measurement errors and outliers are added to test dataset.
Then the root-mean-square error (RMSE) between the estimated power injection and the true power injection is calculated.

\begin{figure}[!t]
	\centering
	\includegraphics[width=0.45\textwidth]{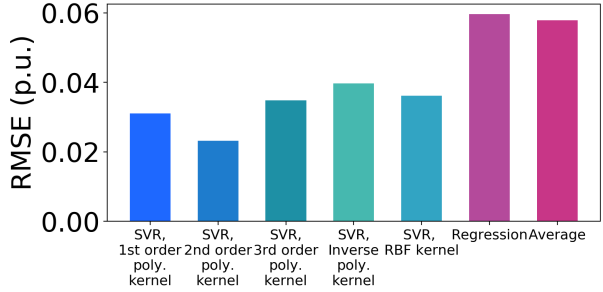}
	\caption{Effectiveness of 2nd order polynomial kernel SVR for forward mapping. The 2nd order polynomial is with the smallest RMSE for forward mapping.}
	\label{fig:g}
\end{figure}

The result for 123-bus case is shown in Fig.~\ref{fig:g}. 
It is clear that the performance of the SVR models are better than the regression model and averaging model.
Among the SVR models with different kernels, the 2nd order polynomial kernel provides the smallest RMSE, supporting the theoretical deduction in Section~\ref{subsec:b}.

\begin{figure}[!t]
	\centering
	\includegraphics[width=0.45\textwidth]{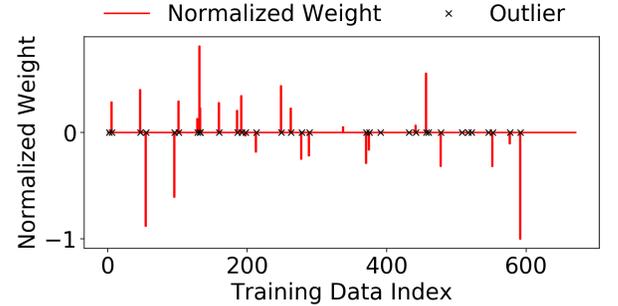}
	\caption{Visualization of support vectors among all training data points.}
	\label{fig:h}
\end{figure}

We also visualize the selection of support vectors among all training data points in Fig.~\ref{fig:h}.
The $x$-axis is the training data point index, and the $y$-axis is the magnitude of the associated dual Lagrangian multipliers.
If a training data point is not a support vector, the dual Lagrangian multiplier is zero.
If a training data point is a support vector, the dual Lagrangian multiplier is nonzero.
In addition, we mark the outlier data points with a black cross.
Fig.~\ref{fig:h} shows that the coefficient is nonzero only when the associated data point is an outlier.

\begin{table}[!t]
\caption{Benchmark of forward mapping}
\label{table:a}
\centering
\begin{tabular}{|c|c|c|c|c|c|c|}
\hline
\multirow{2}{*}{Test Case} & \multicolumn{3}{|c|}{RMSE (p.u.)} & \multicolumn{3}{|c|}{Time Cost (s)}\\
\cline{2-7}
& SVR & Reg & Ave & SVR & Reg & Ave \\
\hline
\hline
8-Bus & 0.023 & 0.060 & 0.058 & 14.5 & 0.0010 & 0.0003\\
\hline
16-Bus & 0.030 & 0.060 & 0.058 & 13.1 & 0.0011.8 & 0.0003\\
\hline
32-Bus & 0.031 & 0.060 & 0.057 & 13.3 & 0.0054 & 0.0007\\
\hline
64-Bus & 0.035 & 0.59 & 0.058 & 14.1 & 0.010 & 0.0004\\
\hline
96-Bus & 0.040 & 0.060 & 0.057 & 14.0 & 0.092 & 0.0005\\
\hline
123-Bus & 0.055 & 0.061 & 0.060 & 15.0 & 0.15 & 0.0005\\
\hline
\specialcell{123-Bus\\ w/ loop} & 0.050 & 0.062 & 0.058 & 15.0 & 0.09 & 0.0006\\
\hline
\end{tabular}
\end{table}

Besides the 123-bus result shown in Fig.~\ref{fig:g}, we also compare the performances on various scales of distribution grids from 8-bus distribution grid to 123-bus distribution grid.
We also test the performances on 123-bus distribution grid with mesh network by manually connecting several buses to mimic some urban systems' weakly meshed structure.
For all of the evaluation, the settings are the same as the 123-bus settings above.
The results are shown in Table~\ref{table:a}.
For all test cases, the RMSEs of SVR model are better than regression model and averaging model.
For the computational time, training SVR models for all cases is within seconds.
This is fast enough for real-time updating.

\subsection{Effectiveness of SVR Model for Inverse Mapping}
\begin{figure}[!t]
	\centering
	\includegraphics[width=0.45\textwidth]{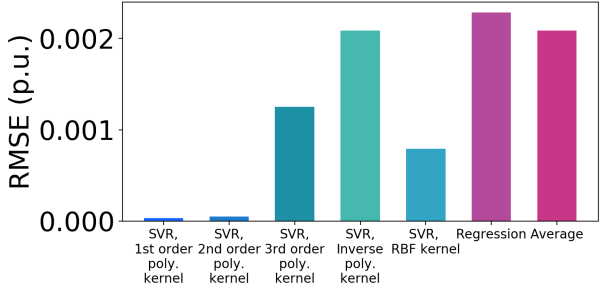}
	\caption{Effectiveness of SVR for Inverse Mapping.}
	\label{fig:i}
\end{figure}

We also test the performance of the SVR model on inverse mapping: from $\pv$, $\qv$ to $|v_i|$ introduced in Section~\ref{subsec:a}.
For inverse mapping, the input of all these models is the active and reactive power injections at all buses, while the output is the voltage magnitude at a particular bus.
Other settings for data preparation and performance evaluation are the same as settings in forward mapping.
The result for 123-bus case is shown in Fig.~\ref{fig:i}.
Differing from the forward mapping, the SVR model with 1st 2nd order polynomial kernel have the best performance.

We evaluate the performances of different models for inverse mapping (from power injections to voltage magnitudes) on various scales of distribution grids from 8-bus to 123-bus distribution grid.
We also test the performance on 123-bus distribution grid with loops.
The results are in Table~\ref{table:b}.
Similar to the results of the forward mapping estimation, the proposed SVR model outperforms other models on all test cases.
The computational times for different models are also similar to the forward mapping estimation.
\begin{table}[!t]
\caption{Benchmark of Inverse Mapping}
\label{table:b}
\centering
\begin{tabular}{|c|c|c|c|c|c|c|}
\hline
\multirow{2}{*}{Test Case} & \multicolumn{3}{|c|}{RMSE ($10^{-3}$p.u.)} & \multicolumn{3}{|c|}{Time Cost (s)}\\
\cline{2-7}
& SVR & Reg & Ave & SVR & Reg & Ave \\
\hline
\hline
8-Bus & 0.20 & 0.66 & 2.1 & 13.0 & 0.0012 & 0.0005\\
\hline
16-Bus & 0.061 & 0.12 & 3.0 & 13.6 & 0.0016 & 0.0006\\
\hline
32-Bus & 0.18 & 0.38 & 2.5 & 13.5 & 0.0068 & 0.0003\\
\hline
64-Bus & 0.60 & 1.4 & 1.8 & 11.7 & 0.026 & 0.0005\\
\hline
96-Bus & 1.1 & 3.2 & 2.6 & 13.2 & 0.057 & 0.0006\\
\hline
123-Bus & 1.9 & 6.0 & 2.8 & 12.0 & 0.095 & 0.0008\\
\hline
\specialcell{123-Bus\\ w/ loop} & 1.9 & 6.5 & 2.8 & 12.2 & 0.10 & 0.0008\\
\hline
\end{tabular}
\end{table}

\subsection{Robustness of the Extrapolation Capability}
\begin{figure}[!t]
	\centering
	\subfloat[No DER, training and testing data in same range.]{\includegraphics[width=0.45\textwidth]{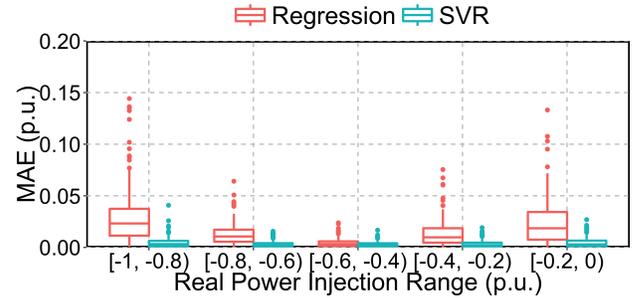}%
	\label{fig:j}}\\
	\subfloat[Testing data with high demands.]{\includegraphics[width=0.45\textwidth]{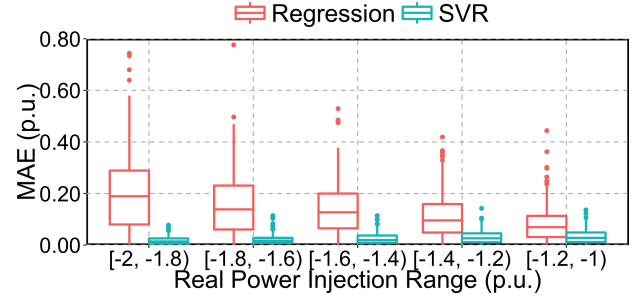}%
	\label{fig:k}}\\
	\subfloat[Testing data with deep DER penetration.]{\includegraphics[width=0.45\textwidth]{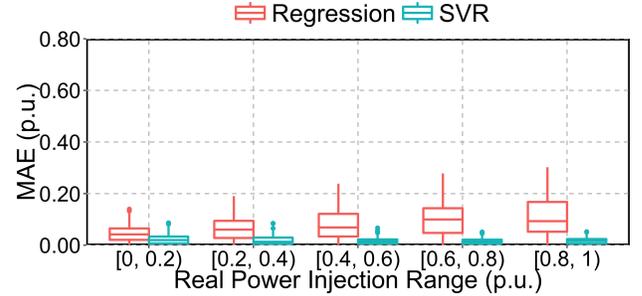}%
	\label{fig:l}}
	\caption{
		Power injection extrapolation performance for testing sets with different ranges. The training set only contains data points with real power injection between $-1$ p.u. and $0$.
	}
	\label{fig:m}
\end{figure}
A power system is a dynamic system, and the load values change significantly over time, especially when it is with different DER penetration levels. 
Therefore, we train the proposed SVR model and the regression model by using a fixed training set, where all the real power injections are within the range of $[-1~\text{p.u.}, 0]$ to obtain the mapping rule from voltage phasors to power injections.
Then, we test the performances of the learned mapping rules in different power injection levels.
We build the same forward mapping model as the previous section, $1\%$ relative measurement error and $2\%$ outliers are added to the training set.

Fig.~\ref{fig:m} demonstrates the performance of the two models, where the mean absolute error (MAE) is used to evaluate the performance.
Fig.~\ref{fig:j} shows the MAEs of estimating the real power injection when the actual power injection range in the test set is the same as the training set.
When the actual power injection is between $-0.6$ p.u. and $-0.4$ p.u. (same range for training and test), the performances of the regression model and the SVR model are both good.
When the actual power injection is around $-1$ p.u. or 0 (slightly different ranges for training and test), the performance of the SVR model is much better than the regression model.
In this case, the performance of the regression model is worse but still acceptable (error is smaller than $0.05$ p.u.).
However, when the range of the testing set is different from the training set, the linear model performs poorly as shown in Fig.~\ref{fig:l}.
In contrast, the performance of the SVR model is much better than the regression model.

In addition to the variation of PV generations in Fig.~\ref{fig:l}, we consider the load variation in Fig.~\ref{fig:k}.
It shows that the SVR model is very robust in such case, and the associated absolute error is always less than $0.1$ p.u in various test cases.
On the other side, the absolute error of regression model could be as high as $0.3$ p.u.
The data for Fig.~\ref{fig:m} is in Table~\ref{table:c}.
\begin{table}[!t]
\caption{Forward Mapping: Mapping from Voltage Phasors to Power Injection}
\label{table:c}
\centering
\begin{tabular}{|c|c|c||c|c|c|}
\hline
& \multicolumn{2}{|c||}{MAE} & & \multicolumn{2}{|c|}{MAE}\\
\hline
Range & Reg & SVR & Range & Reg & SVR\\
\hline
[-2.0, -1.8) & 0.2036 & 0.0185 & [-0.4, -0.2) & 0.0135 & 0.0035\\
\hline
[-1.8, -1.6) & 0.1625 & 0.0220 & [-0.2, 0.0) & 0.0246 & 0.0048\\
\hline
[-1.6, -1.4) & 0.1448 & 0.0258 & [0.0, 0.2) & 0.0467 & 0.0229\\
\hline
[-1.4, -1.2) & 0.1144 & 0.0320 & [0.2, 0.4) & 0.0631 & 0.0204\\
\hline
[-1.2, -1.0) & 0.0833 & 0.0325 & [0.4, 0.6) & 0.0792 & 0.0166\\
\hline
[-1.0, -0.8) & 0.0292 & 0.0045 & [0.6, 0.8) & 0.1019 & 0.0155\\
\hline
[-0.8, -0.6) & 0.0129 & 0.0031 & [0.8, 1.0] & 0.1102 & 0.0174\\
\hline
[-0.6, -0.4) & 0.0045 & 0.0030 &\multicolumn{3}{|c|}{}\\
\hline
\end{tabular}
\end{table}

We also test the extrapolation ability of the SVR model for the inverse mapping from power injections to the voltage magnitude at a certain bus of the distribution grid.
Similar to use case one, we investigate the performance of the proposed model in different power injection levels.
Table~\ref{table:d} presents the detailed results for inverse mapping estimation.
When the training data and testing data are in the same range, the performances of the SVR model is better than the linear regression model, while the error of the regression model is still relatively small, e.g., MAE is less than $0.002$ p.u.
However, when the power injection range of the testing set is different from the range in the training set, the performance of the regression model degrades significantly, while the SVR model retains good performance.
\begin{table}[!t]
\caption{Inverse Mapping: Mapping from Power Injections to Voltage Magnitude}
\label{table:d}
\centering
\begin{tabular}{|c|c|c|c|}
\hline
\specialcell{Real Power\\Injection Range} &\specialcell{Voltage\\Magnitude Range} & \specialcell{MAE of\\ Regression} & \specialcell{MAE of\\ SVR} \\
\hline
\hline
\multirow{4}{*}{[-2.0, -1.0)} & [0.75, 0.90) & 0.0180 & 0.0028\\
\cline{2-4}
& [0.90, 0.95) & 0.0170 & 0.0016\\
\cline{2-4}
& [0.95, 1.00) & 0.0167 & 0.0007\\
\cline{2-4}
& [1.00, 1.10) & 0.0157 & 0.0004\\
\hline
\hline
\multirow{4}{*}{[-1.0, 0.0)} & [0.80, 0.95) & 0.0007 & 0.00002\\
\cline{2-4}
& [0.95, 1.00) & 0.0003 & 0.00002\\
\cline{2-4}
& [1.00, 1.05) & 0.0003 & 0.00003\\
\cline{2-4}
& [1.05, 1.10) & 0.0004 & 0.00003\\
\hline
\hline
\multirow{4}{*}{[0.0, 1.0)} & [0.90, 1.00) & 0.0165 & 0.0005\\
\cline{2-4}
& [1.00, 1.05) & 0.0155 & 0.0004\\
\cline{2-4}
& [1.05, 1.10) & 0.0143 & 0.0011\\
\cline{2-4}
& [1.10, 1.20) & 0.0126 & 0.0015\\
\hline
\end{tabular}
\end{table}

\subsection{Robustness to Outliers}
We test the robustness of the proposed SVR model to outliers in training data.
In this test, no random measurement errors are added to training data. 
We modify the percentage of outliers from $0$ to $8\%$ in training set for \textit{all} direct measurements.
Six weeks' data is used for training and validation, while another three weeks' data is used for testing.
We use mean squared error (MSE) to evaluate the performances.

\begin{figure}[!t]
	\centering
	\includegraphics[width=0.45\textwidth]{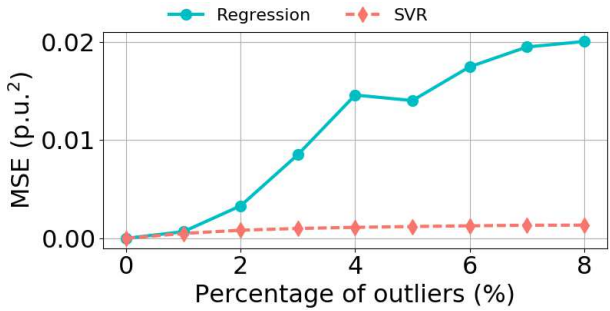}
	\caption{Performances of regression and SVR with different level of outliers in training data.}
	\label{fig:n}
\end{figure}

For forward mapping, the performance for 123-bus case is illustrated in Fig.~\ref{fig:n}.
When there are no outliers, both of the regression model and the proposed SVR model work well.
However, the MSE of the regression method increases fast even if there are only 2\% outliers in training data, while the performance of the SVR method is robust enough for 8\% outliers in training data.

\subsection{Flexibility for Active Controllers}
\begin{figure}[!t]
	\centering
	\includegraphics[width=0.45\textwidth]{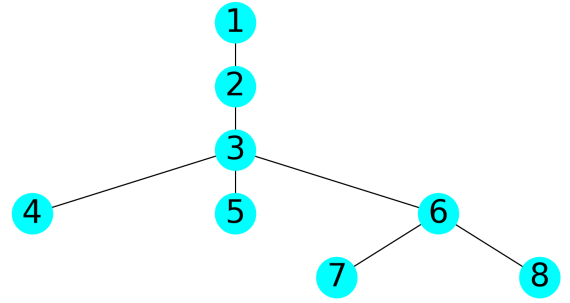}
	\caption{Illustration of an 8-bus distribution grid with full measurements.}
	\label{fig:o}
\end{figure}

We test the performance of the proposed SVR model when there exists active controllers in the system.
In particular, we add a droop controller at bus 7 in an 8-bus distribution grid, which is shown in Fig.~\ref{fig:o}.

The control signals are the voltage magnitude at bus 7, or average voltage magnitude at bus $4, 5, 7$ and $8$, which stabilizes the voltage of a single bus or the voltages at the leaves of the network, correspondingly.
The voltage droop coefficient is $10$, which means $0.05$ p.u. in voltage change introduces $0.5$ p.u. in controlled reactive power bank output change.
In this case, we build the forward mapping from voltages at all buses to the reactive power injection at bus 7, which is affected by the droop controller.
No random measurement errors and outliers are added to training data.
Other settings of the training and testing for regression model and the SVR model are similar with previous settings.

\begin{figure}[!t]
	\centering
	\includegraphics[width=0.45\textwidth]{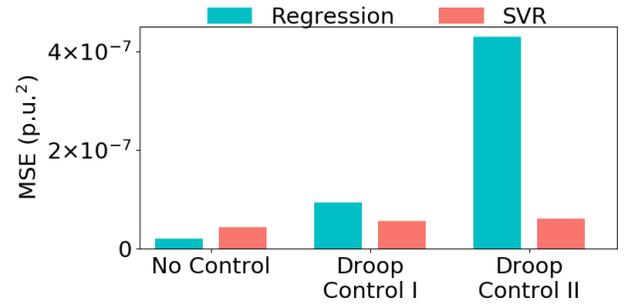}
	\caption{Performances of regression and SVR with different droop controllers in distribution grids. Controller I adjusts the reactive power injection at bus 7 based on the voltage magnitude at bus 7. Controller II adjusts the reactive power injection at bus 7 based on the mean of the voltage magnitudes at bus 4, 5, 7, and 8.}
	\label{fig:p}
\end{figure}

Fig.~\ref{fig:p} shows the performance of the SVR method when there are unmodeled active controllers in distribution grids.
Given a smaller size of network, and without measurement errors and outliers, the performance difference solely depends on the existence of the unmodeled controllers.
When there is no active controller, the learned mapping rule for both regression and SVR is accurate.
However, with the unmodeled active controller, the performance of the regression is much worse than SVR, especially for the controller II, which adjusts the reactive power injection at bus 7 based on the mean of the voltage magnitudes at bus 4, 5, 7, and 8.

\begin{figure}[!t]
	\centering
	\includegraphics[width=0.45\textwidth]{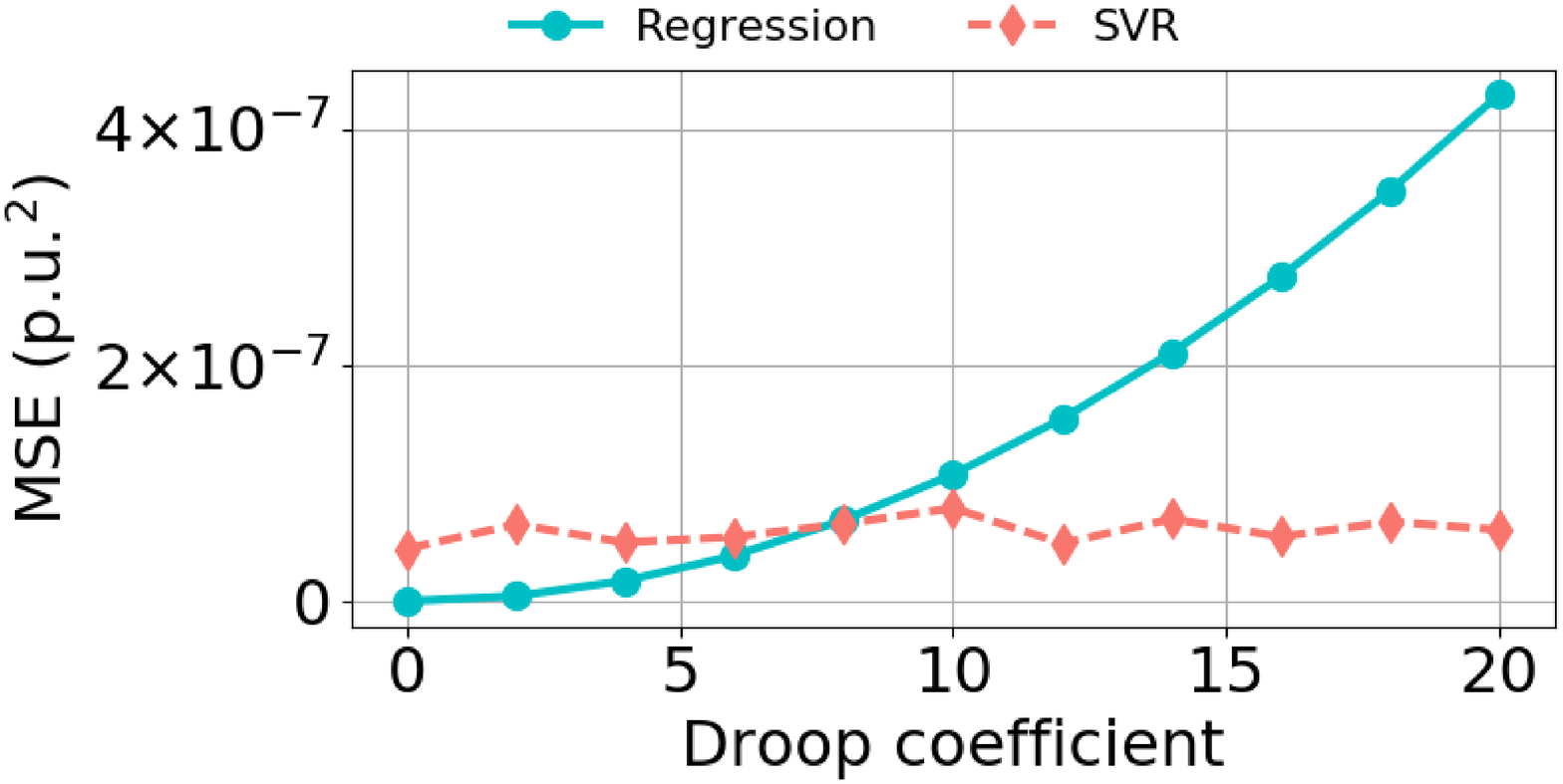}
	\caption{Performances of regression and SVR with different level of droop coefficients.}
	\label{fig:q}
\end{figure}

We further evaluate the performance for both SVR and regression with different voltage droop coefficient.
The results are shown in Fig.~\ref{fig:q}.
In the evaluation, we choose to use the droop controller to adjust the reactive power injection at bus 7, with the input of the average voltage magnitude at bus $4, 5, 7$ and $8$.
The performance of the SVR is very robust against the increase of the droop coefficient, which proves that the SVR model can learn the controller's mechanisms while the parameter-learning approach cannot.

\subsection{Robustness to Partial Observations}
\begin{figure}[!t]
	\centering
	\includegraphics[width=0.45\textwidth]{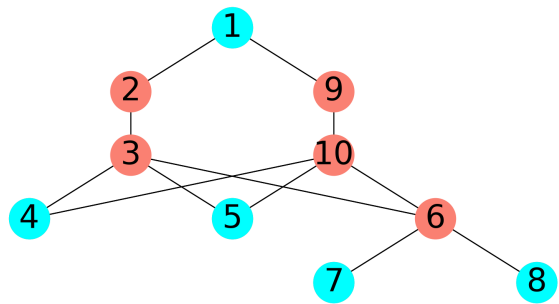}
	\caption{A 10-bus distribution grid with mesh network. Measurements are only available at buses with green color.}
	\label{fig:r}
\end{figure}

We conduct the performance test of the proposed SVR method when we have partial available measurements in a distribution grid.
In this section, we choose a 10-bus test distribution grid with mesh network, while bus 1 is the slack bus, shown in Fig.~\ref{fig:r}.
We only have the measurements of voltages and power injections at bus 1, 4, 5, 7, 8, which are colored green in Fig.~\ref{fig:r}.

For this testing, no random measurement errors or outliers are added in training set.
We consider two setups.
One setup is that there are no net power injections at the buses without measurement, bus 2, 3, 6, 9, 10.
In this case, we may introduce an an equivalent admittance matrix which represents a fully connected graph among \emph{active} buses which have non-zero power injections by Kron reduction of the admittance matrix~\cite{dorfler2013kron}.
Another setup is that there are net power injections at the hidden buses.
The net power injection at hidden buses could be private controllers such as reactive power banks or energy losses.
We model the net power injection at hidden buses as energy losses which are proportional to the energy consumptions at leaves nodes.
No random measurement errors and outliers are added to the training data.
Other settings of the training and testing are similar with previous settings.

\begin{figure}[!t]
	\centering
	\includegraphics[width=0.45\textwidth]{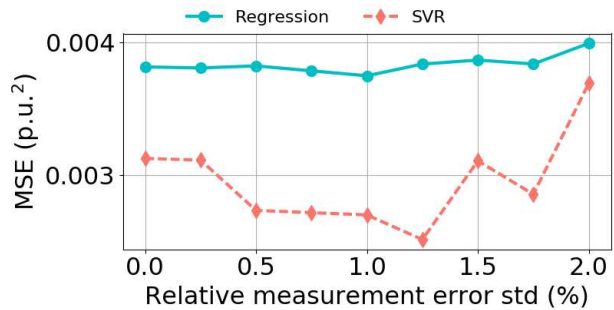}
	\caption{Performances of regression and SVR for partial observations in distribution grids with different levels of measurement errors.}
	\label{fig:s}
\end{figure}

We further add random measurement errors with different levels of standard deviations to training set, and test the robustness of the SVR model when there are measurement errors.
The result is shown in Fig.~\ref{fig:s}.
It is shown that no matter what the relative measurement error is, the SVR's performance for partially available data is better than regression model, implying the better modeling flexibility and robustness.

\section{Conclusion}\label{sec:e}
With deep DER penetration in distribution grids, proper monitoring with current sensor capability is needed. 
As the topology, parameters, and active controller information are usually insufficient in some primary distribution grids and many secondary distribution grids, it is hard to apply the transmission grid monitoring tools directly. 
In this paper, we propose a data-driven approach to recover the mapping rules of power flow equations in distribution grids thanks to the flexibility of kernel trick design in support vector regression. 
We prove that the data-driven SVR method can match the traditional physical law-based regression method exactly when there are perfect measurements. 
This property resolves a typical drawback of data fitting methods for overfitting. 
Numerical results show that our method is robust when there are outliers in historical measurements and when only partial measurements are available. 
These advantages make our proposed method promising to apply in distribution grids and serves as the basis for other power flow-based for applications, such as state estimation and optimal power flow.

\bibliographystyle{IEEEtran}
\bibliography{bibTex}
\end{document}